\documentclass[a4paper,twoside]{article}
\newcommand{\affil}[1]{$^{\rm #1}$}
\usepackage[authoryear]{natbib}
\bibpunct{(}{)}{;}{a}{}{,}
\usepackage{graphicx}
\date{} 
\newcommand{\HI}{H{\sc i} }
\usepackage{aas_macros}
\title{\large\bf\flushleft The Characterised Noise \HI source finder: Detecting \HI galaxies using a novel implementation of matched filtering}
\author{\parbox{\textwidth}{\flushleft
\vspace{-0.5cm}
{\it R. Jurek\affil{A,B}}\\
\vspace{0.4cm}
{\small \affil{A}\,CSIRO Astronomy \& Space Sciences, Australia Telescope National Facility, P.O. Box 76, Epping NSW 1710, Australia.}\\
{\small \affil{B}\,Email: Russell.Jurek@CSIRO.au}}}
\begin{document}
\twocolumn[
\begin{changemargin}{.8cm}{.5cm}
\begin{minipage}{.9\textwidth}
\vspace{-1cm}
\maketitle
%
%
\small{\bf Abstract:}
The spectral line datacubes obtained from the Square Kilometre Array (SKA) and its precursors, such as the Australian SKA Pathfinder (ASKAP), will be sufficiently large to necessitate automated detection and parametrisation of sources. Matched filtering is widely acknowledged as the best possible method for the automated detection of sources. This paper presents the Characterised Noise \HI ({\sc CNHI}) source finder, which employs a novel implementation of matched filtering. This implementation is optimised for the 3-D nature of the planned Wide-field ASKAP Legacy L-band All-sky Blind surveY's (WALLABY) \HI spectral line observations. The {\sc CNHI} source finder also employs a novel sparse representation of 3-D objects, with a high compression rate, to implement Lutz one-pass algorithm on datacubes that are too large to process in a single pass.

WALLABY will use ASKAP's phenomenal 30 square degree field of view to image $\sim 70\%$ of the sky. It is expected that WALLABY will find 500 000 \HI galaxies out to $z \sim 0.2$.


\medskip
\medskip
\end{minipage}
\end{changemargin}
]
\small

\section{Introduction}
The Wide-field ASKAP Legacy L-band All-sky Blind surveY (WALLABY)\footnote{Principal Investigators: Baerbel Koribalski and Lister Staveley-Smith. See www.atnf.csiro.au/research/WALLABY for more details about the survey.} (\citet{Koribalski_2009}; Koribalski, B., Staveley-Smith, L. et~al., in preparation) is an ambitious project that aims to detect neutral hydrogen to a redshift of $z \sim 0.26$, across $\sim 70\%$ of the sky. It is one of the two top ranked projects that will be carried out using the Australian SKA Pathfinder (ASKAP). WALLABY is possible because of ASKAP's unprecedented $\sim 30$ sq. degree field-of-view, which is achieved using Phased Array Feeds (PAFs). WALLABY will use all 36 of ASKAP's antennae, but due to limitations on computing resources will only process the inner 30 antennae (with a maximum baseline of 2km) to image the sky with a 30$^{\prime\prime}$ synthesised beam and produce datacubes with voxels\footnote{Voxels are often referred to as pixels when discussing a single channel of a datacube. Technically though these `pixels' are still voxels. For this reason the term voxel is used throughout instead of pixel to aid consistency.} that project to $\sim$10$^{\prime\prime}$ on the sky. The high spatial resolution is complemented by an anticipated spectral resolution of 3.86 km s$^{-1}$. ASKAP spectral datacubes will therefore cover a large area of the sky to high resolution, which results in very large datacubes containing at least 2048 x 2048 x 16 384 voxels. WALLABY will consist of $\sim$ 1200 of these large datacubes. The size and number of these datacubes renders manual source finding unfeasible. The performance of the automatic source finder used by WALLABY will determine how many (\HI) galaxies are found by WALLABY.

The majority of source finders in existence use intensity thresholding to find sources. {\sc SExtractor} \citep{1996A&AS..117..393B}, {\sc SFind} \citep{2002AJ....123.1086H} and {\sc Duchamp} \citep{2008glv..book..343W,Duchamp2} are good examples of source finders based on intensity thresholding. Conceptually, intensity threshold source finders check every pixel (voxel) in an image (datacube) to see if the pixel (voxel) value is sufficiently extreme that it's unlikely to be noise. Once all of the source pixels (voxels) have been identified, they are combined into objects. The various intensity threshold source finders differ in how they estimate the noise, set a threshold for identifying source pixels (voxels), pre-process the image (datacube) to improve the source finder results and the manner in which they create objects from source pixels (voxels). All intensity threshold source finders share an inherent limitation though. 

\begin{figure}[ht]
\begin{center}
\includegraphics[width=0.98\linewidth]{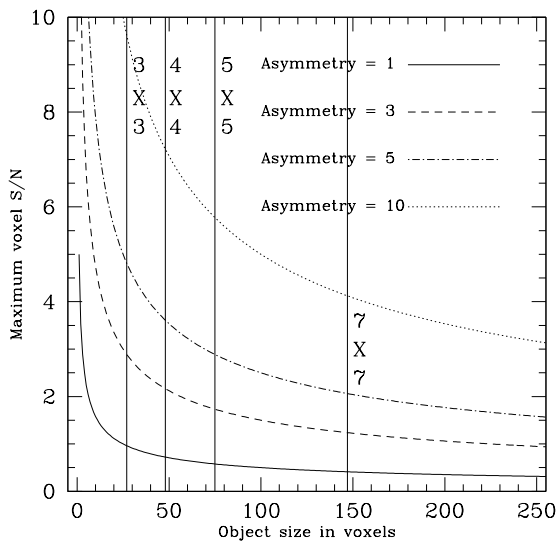}
\caption{The maximum voxel $S/N$ of an object (with an integrated $S/N = 5$) plotted against the number of voxels comprising the object, n. The various lines correspond to different asymmetries in the distribution of the voxel's flux over the n voxels. The vertical lines (labelled) denote the minimum size of a point source extending over three channels and occupying 3x3, 4x4, 5x5 or 7x7 voxels in every channel.}
\label{fig:VoxelSN}
\end{center}
\end{figure}

Consider an arbitrary source in a spectral datacube. Improved spatial and spectral resolutions result in the source occupying more voxels in the datacube. Dispersing the source's signal over more voxels means that it contributes less to the flux value of each voxel that it occupies. This makes it harder for an intensity threshold method to detect the source. Using a simple model this effect is illustrated in Figure \ref{fig:VoxelSN}, where the maximum voxel $S/N$ of an object with an integrated $S/N$ of 5 is plotted for various asymmetries. The maximum voxel $S/N$ is calculated to be $S/N_{integrated} \times \beta / \sqrt{n}$, where $\beta$ describes the asymmetry of the object's flux distribution and $n$ is the number of voxels. 

By overlaying the minimum expected size (in voxels) of WALLABY sources on Figure \ref{fig:VoxelSN}, we can assess the impact of this inherent limitation on WALLABY. The neutral hydrogen detected in emission is warm, which gives it an intrinsic amount of dispersion. We will assume that any real WALLABY source extends over at least 3 channels. We will also assume that in every channel, a source occupies at least 3x3 voxels for a 30$^{\prime\prime}$ synthesised beam and 10$^{\prime\prime}$ voxels). Galaxies rarely lie at the middle of a voxel though, so a more realistic minimum is a grid of 4x4 or 5x5 voxels in every channel. It is expected that most galaxies detected by WALLABY will be unresolved or at most marginally resolved (Duffy, A., Meyer, M. \& Staveley-Smith, L. 2011, in preparation), so we also consider a 7x7 grid of voxels in every channel to account for marginally resolved, off-centre galaxies. After multiplying by the minimum number of channels to obtain the expected minimum size of WALLABY galaxies in voxels, the minimum size for these different grids is overlaid in Figure \ref{fig:VoxelSN}. This demonstrates that off-centre and/or marginally resolved galaxies will be difficult to detect with a basic implementation of an intensity thresholding source finder, unless the flux is asymmetrically distributed. Figure \ref{fig:VoxelSN} also illustrates that this effect is amplified in 3-D datasets such as future WALLABY datacubes. For example, in a 2-D image the 7x7 and 5x5 vertical lines would approximately lie at the position of the 4x4 and 3x3 vertical lines. 

This inherent limitation is compounded further by using a size-based rejection criterion to weed out false detections. If you only detect a few, unconnected voxels the source will be flagged as a false positive by a size-based rejection criterion. Off-centre and/or marginally resolved galaxies that are detected because of asymmetric flux distributions are most likely to be detected in the form of a few, unconnected voxels. This effect will also show up as an enhanced fracture rate of extended, well resolved sources.

The inherent limitation of intensity thresholding based source finders can be offset by using more aggressive intensity thresholds (i.e., lower intensity thresholds), but it often results in many false (source) detections. The solution to this problem is to run the source finder on the datacube multiple times with the datacube smoothed to a different scale each time. This is however a very inefficient solution to the problem. {\sc Duchamp} provides multiple options for dealing with this inherent limitation: a secondary `growth' threshold, smoothing and a 3-dimensional wavelet reconstruction of the dataset.

As part of its design study, WALLABY has investigated novel methods of source detection as an alternative to multiple passes of an intensity threshold source finder. The goal of this investigation was to develop source detection methods that are optimised for large datacubes with high spatial and spectral resolution. The Characterised Noise \HI ({\sc CNHI}) source finder that I present here is one of the novel source detection methods that have been developed.

The rest of the paper is structured as follows. The conceptual framework for the {\sc CNHI} source finder is presented in Section \ref{section:concept}. The inherent limitations of the {\sc CNHI} source finder are then discussed in Section \ref{section:limits}. Next, the current implementation of the {\sc CNHI} source finder is presented in Section \ref{section:implement}. Finally, some example results are discussed in Section \ref{section:examples} before finishing with a summary.

\section{Conceptual framework}
\label{section:concept}
The {\sc CNHI} source finder is based on three concepts. The first concept is that WALLABY spectral datacubes can be treated as a bundle of \HI spectra, rather than a collection of voxels. The next concept is that contiguous blocks of voxels should be tested to see if they're a source, rather than individual voxels. The final concept is that a `source' is detected by looking for a region in a datacube that doesn't look like noise. This is the inverse of most source finders which identify sources based on some idea of what a source looks like. In the rest of this section these concepts are explained in detail.

The first part of the {\sc CNHI} source finder conceptual framework is treating a WALLABY spectral datacube as a bundle of \HI spectra, which is akin to Integral Field Unit (IFU) observations. Each position on the sky has its own spectrum. Each spectrum in this datacube is correlated to some degree however with the neighbouring spectra. The ASKAP beam will determine the degree of correlation between each spectrum and its neighbouring spectra. As explained later the correlation between spectra should not be a problem for the {\sc CNHI} source finder. This conceptual view of a WALLABY datacube is very amenable to parallelisation. 

The second component of the {\sc CNHI} source finder conceptual framework is to test contiguous blocks of voxels instead of individual voxels. This concept is designed to take advantage of WALLABY datacubes having sufficiently high velocity resolution to reasonably resolve most galaxies in frequency. As discussed above, high resolution data is problematic for source finding methods that analyse individual voxels. The high resolution is however advantageous when testing contiguous blocks of voxels. A single voxel with a flux value that is one standard deviation above the mean is not significant. Ten contiguous voxels that are all one standard deviation above the mean are, because it's improbable that this will happen by chance in a spectrum with negligible correlation. If we can test whether a contiguous block of voxels in a spectrum is likely to be source, then we are searching for sources in a way that is benefited by the high velocity resolution of WALLABY datacubes. 

Testing whether contiguous blocks of voxels are sources provides additional information compared to testing individual voxels. It is a reasonable expectation that the test region that best fits the position and velocity width of a source in a given \HI spectrum is the most significant test region. If the test region is too small, then it should be less significant than a larger region that is also made up of source voxels. If the test region is too large, then the test region contains both source and pure noise voxels, which should result in a less significant test region. Identifying the position and width that results in the most significant test region, therefore provides an estimate of the source position and velocity width.

The final concept of the {\sc CNHI} source finder is to find sources by looking for regions in a WALLABY datacube that do not look like noise. Looking for regions that do not look like noise is a novel way to implement matched filtering. Rather than using many, many filters that each describe a different type of source, we can use a single filter that looks like noise. This works because we can safely assume that the presence of a source, is what causes a contiguous region in our datacube to not look like noise. The key is to use the noise distribution as the filter, because it is relatively stable, even though individual realisations of the noise vary.

How do we use the noise distribution as a filter to look for regions that aren't pure noise? Due to the large size and high resolution of WALLABY datacubes they are expected to be sparsely populated by sources. This means that an arbitrary \HI spectrum will be dominated by noise. If we select a test region of contiguous voxels, then we can use the rest of the \HI spectrum as an example of noise. A comparative statistical test such as the Kolmogorov-Smirnov test \citep{1979ats..book.....K} can then determine the probability that the test region and the rest of the spectrum, which is noise dominated, come from the same distribution of voxel fluxes. In other words, using a comparative statistical test we can identify if a test region looks like noise. This implementation can easily be adapted to 2-D images and spectra.

The {\sc CNHI} source finder uses the Kuiper test \citep{Kuiper1960} to compare the voxel flux distribution of a test region to the rest of the LoS spectrum, and identify regions with non-noise voxel flux distributions. The Kuiper test is a variant of the Kolmogorov-Smirnov test that is cyclically-invariant. The Kolmogorov-Smirnov test is most sensitive to differences in the distributions about the medians of the distributions. The Kuiper test by contrast is equally sensitive to differences in the two distributions throughout their entire range. 

\section{Inherent limitations}
\label{section:limits}
There are two inherent limitations of the {\sc CNHI} source finder. These limitations are tied to the conceptual framework. The first limitation is that the comparison of the test region with the rest of the \HI spectrum relies on a noise dominated \HI spectrum. The other limitation is that a test region needs to be sufficiently large for the Kuiper test to produce reliable results.

The use of the Kuiper test to determine if a region in a \HI spectrum is noise-like, is predicated upon the assumption that the voxel flux values of the rest of the \HI spectrum are noise. It is expected that the assumption of noise dominated \HI spectra is valid for properly calibrated, flagged and continuum subtracted WALLABY datacubes with no significant baseline structure. To illustrate the sparsity of WALLABY datacubes, a typical WALLABY datacube is only 0.6\% source (measured in voxels) if 500 000 sources distributed across 1 200 datacubes have a typical size of 1 000 000 voxels. 

For datacubes that are not well behaved or sufficiently sparse (e.g., baseline structure or failed bandpass calibration), this can be dealt with by comparing a test region in a \HI spectrum to the subset of the remaining \HI spectrum in its immediate vicinity. The Kuiper test will be less sensitive, but this is offset by making a valid statistical comparison. Fourier analysis, polynomial fitting and existing baseline structure removal techniques (such as those implemented in {\sc Duchamp}) can also be used in combination with this approach, or as an alternative.

The validity of a Kuiper test is described by the $Q$ parameter. For two samples containing $n_1$ and $n_2$ values, the $Q$ parameter is $n_1 \times n_2 / (n_1 + n_2)$. This is the same $Q$ parameter that describes the validity of the Kolmogorov-Smirnov test. For both the Kuiper and Kolmogorov-Smirnov tests it is accepted that the test results are valid for $Q \geq 4$. The test region in a \HI spectrum needs to be sufficiently large to satisfy $Q \geq 4$, otherwise the Kuiper test results are increasingly spurious for increasingly smaller test regions. Setting $n = n_1 + n_2$, $Q = 4$ and letting $m$ be the minimum size of a test region, then we can solve for $m$. The minimum size of a test region is $m = ( n - \sqrt{n^2 - 16n} ) / 2$. 

For a WALLABY datacube the minimum size of a test region is 4 channels. This matches the minimum expected channel width of WALLABY \HI galaxies. The use of the {\sc CNHI} source finder on other datacubes needs to consider the minimum size of a test region. For reference the minimum test region size is 4 channels for $n \geq 40$, and the Kuiper test can not achieve $Q \geq 4$ for $n \leq 15$. 

\section{Current implementation}
\label{section:implement}
\begin{figure*}[htbp]
\begin{center}
\includegraphics[width=0.99\textwidth]{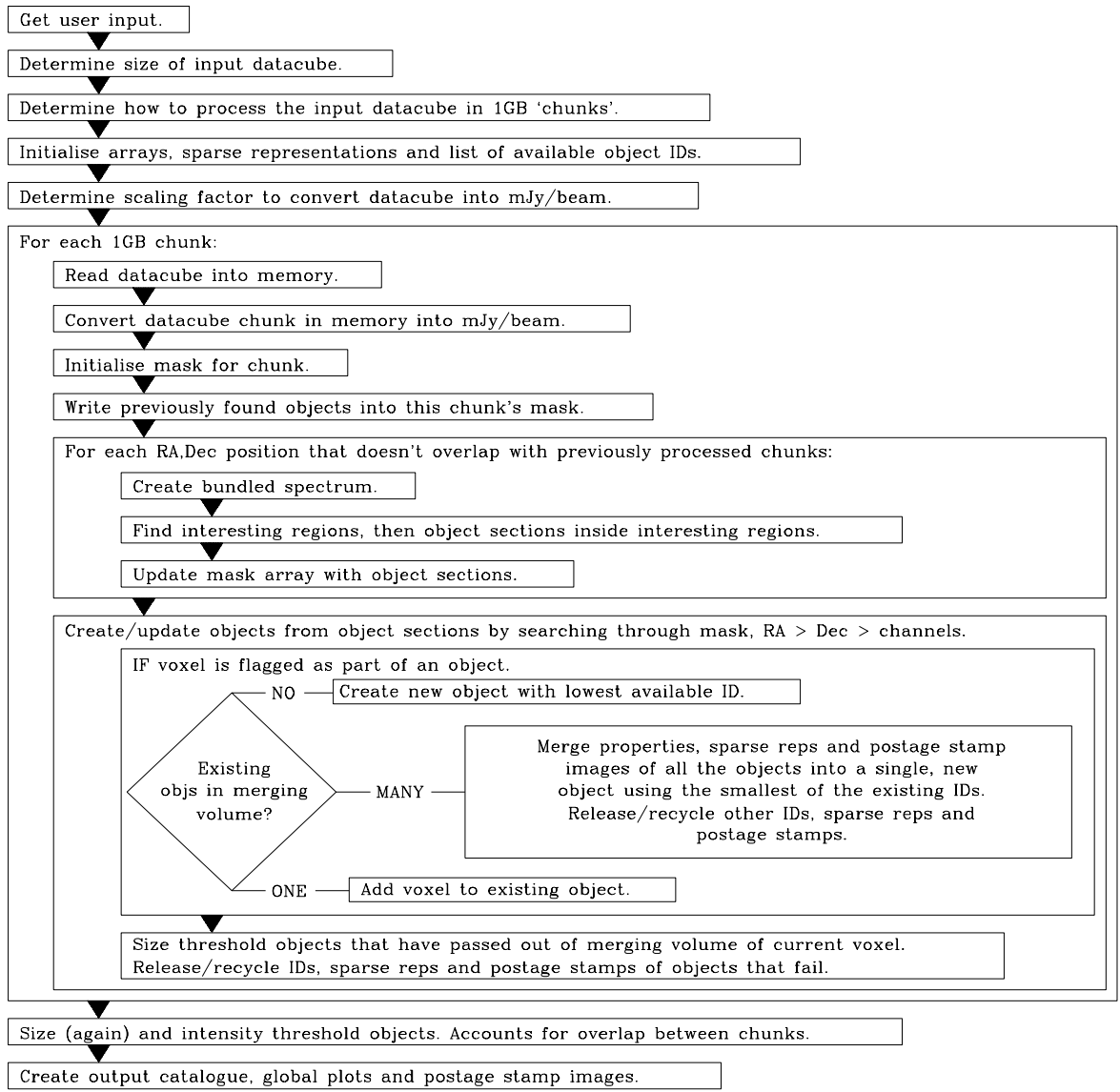}
\caption{The algorithm describing the current implementation of the {\sc CNHI} source finder is shown here as a flow diagram.}
\label{fig:flow}
\end{center}

\end{figure*}

The current implementation of the {\sc CNHI} source finder works in the following manner. A user calls the {\sc CNHI} source finder from the command line with a list of input parameters. The software then figures out how many chunks to split the input file into, such that each chunk is at most 1GB in size. For each chunk, the software creates a bundled \HI spectrum for each position on the sky and uses the Kuiper test to find object sections in the bundled spectrum. Once all of the bundled \HI spectra have been searched for object sections, the software creates objects out of them using a variant of Lutz one pass algorithm \citep{1980CompJ..23..262L}. The list of objects, their properties and postage stamp images are external to the chunks and new objects are added as each chunk is processed. Once all of the chunks have been processed, the final list of objects is tested against the user specified rejection criterion. The objects that remain are then output to a catalogue, added to global moment 0 and position-velocity plots and postage stamp images (including integrated spectra) generated. A flow diagram of the current implementation is presented in Figure \ref{fig:flow}. The following subsections provide more detail about the {\sc CNHI} input and output, the bundling of multiple \HI spectra, finding object sections in a bundled \HI spectrum and the creation of objects from object sections.

\subsection{Inputs and outputs}
\label{section:input}
User input consists of the following parameters:
\begin{enumerate}
\item{Output code: The output catalogue and plots are created with names of OutputCode\_obj.cat, OutputCode\_plots.ps and OutputCode\_spectra.ps.}
\item{File name: The path to and name of the .fits file that {\sc CNHI} will search for sources.}
\item{Pre-threshold: The probability threshold used to identify interesting regions that are likely to contain a source\footnote{Note that a higher value digs deeper into the noise, and is equivalent to using a lower intensity threshold.}. (Dimensionless quantity.)}
\item{Threshold: The probability threshold used to identify sources within interesting regions\footnotemark[\value{footnote}]. (Dimensionless quantity.)}
\item{Minimum bounding box (3 values): The size criterion applied to an object's bounding box if it is to be retained.}
\item{Bounding box filling factor: An object composed of fewer voxels than the minimum bounding box multiplied by the filling factor is rejected.}
\item{Pseudo total intensity threshold: Objects with a pseudo total intensity less than this threshold are rejected.}
\item{Merging distances (3 values): Objects which are separated by this many voxels or fewer in any dimension are merged into a single object.}
\item{Maximum scale: The maximum size of a test region in a \HI spectrum. Specified in number of channels.}
\end{enumerate} 

The {\sc CNHI} source finder output consists of a catalogue, a global moment 0 map, a global position-velocity diagram for both RA and Dec, and postage stamp images of each object (including an integrated spectrum). The {\sc CNHI} source finder catalogue contains the following information for each object:
\begin{enumerate}
\item{ID: A numerical ID assigned to each object.}
\item{Voxel count: The number of voxels that constitute the source.}
\item{Pixel RA: The mean RA of the object's voxels.}
\item{Pixel Dec: The mean Dec of the object's voxels.}
\item{Pixel channel: The mean channel of the object's voxels.}
\item{Intensity RA: The flux weighted mean RA of the object's voxels.}
\item{Intensity Dec: The flux weighted mean Dec of the object's voxels.}
\item{Intensity channel: The flux weighted mean channel of the object's voxels.}
\item{Voxel limits (6 values): The minimum and maximum RA, Dec and channel of the object.}
\item{Voxel flux statistics (5 values): The sum, mean, minimum, maximum and standard deviation of the object's voxel flux values.}
\item{Two sets of $W_{20}$ and $W_{50}$ measurements.}
\item{Sparse representation: A sequence of values that describes a sparse representation of the object's 3-D bit mask in the datacubes voxel co-ordinates.}
\end{enumerate}

The $W_{20}$ and $W_{50}$ measurements are measured in two ways. The first set of $W_{20}$ and $W_{50}$ measurements uses the same method as {\sc Duchamp}. First, the global maximum of the object's integrated spectrum is determined. Next, starting from each end of the integrated spectrum the first channel with a flux greater than or equal to 20\% and 50\% of the maximum are identified and used to measure $W_{20}$ and $W_{50}$. The second set of width measurements was developed as part of the source finder framework that is used to implement {\sc CNHI}. First, the total flux of the integrated spectrum is measured. The cumulative frequency distribution (cfd) of the total flux is then constructed as a function of channel number for the integrated spectrum. The inner 92.7\% and 76.1\%, which corresponds to the conventional $W_{20}$ \& $W_{50}$ for a gaussian profile, are then used to determine $W_{20}$ and $W_{50}$. The cfd will oscillate at the edges of the integrated spectrum because of noise. For this reason the `inner' values are defined (starting from the left edge of the integrated spectrum) to start at the channels where the cfd never again dips below 0.0364 ($W_{20}$) and 0.1195 ($W_{50}$), and ends at the channels where the cfd first rises above 0.9636 ($W_{20}$) and 0.8805 ($W_{50}$). This approach to measuring $W_{20}$ and $W_{50}$ has a consistent physical meaning across all possible spectral profiles and `in principle' averages out noise.

\subsection{Creating bundled \HI spectra}
\label{section:bundling}
For each \HI spectrum the {\sc CNHI} source finder bundles together this \HI spectrum and the neighbouring \HI spectra. This bundled \HI spectrum is searched for objects. A bundled spectrum is created by weighting the sum of the \HI spectrum and it's neighbouring \HI spectra using the point spread function. 

Searching for object sections in a bundled spectrum has two advantages. The first advantage is that it slightly improves the $S/N$ without blurring the edges of sources in velocity space or creating correlation in the bundled \HI spectrum. The second, more important advantage is the improved detection of the outer edges of a source. Bundling the \HI spectra couples the brighter flux in an object's inner region to the fainter flux at the edge of the object. This improves the chance of detecting the fainter, outer regions of the object. The amount of improvement varies non-linearly with source morphology, user input and datacube, so no attempt is made here to predict the amount of improvement.

\subsection{Finding object sections}
\label{section:finding}
The Kuiper test is used to find object sections within a bundled \HI spectrum. This is implemented as a four step process. In the first step, the Kuiper test is used to identify interesting regions that are likely to contain an object section. The next step is to reduce the list of interesting regions to a unique set. This is achieved by finding all of the interesting regions that overlap, and only keeping the most significant interesting region. The most significant interesting region is judged to be the region with the lowest probability of being noise according to the Kuiper test. The purpose of these first two steps is to efficiently reduce the bundled spectrum to a manageable subset. Third, the Kuiper test is applied to every position on every relevant scale to find object sections within these interesting regions. The final step is to reduce the object sections to a unique set. This is achieved in the same way that interesting regions are reduced to a unique set.

Interesting regions are found by applying the Kuiper test to test regions in the bundled \HI spectrum and comparing the result to a user defined pre-threshold. Interesting regions are found efficiently by Nyquist sampling both the scales of interest and positions along the bundled \HI spectrum. Starting with the largest scale of interest (user specified) and the beginning of the bundled \HI spectrum, the Kuiper test is applied and then the test region is advanced half of the scale length. Once the entire bundled spectrum has been tested on this scale, the scale is halved and the process is repeated. This is repeated until the minimum scale has been processed.

Interesting regions narrow down the location and scale of object sections. Only positions within the interesting region and scales ranging from half as small as the interesting region up to the size of the interesting region, need to be investigated. If other positions or scales were a better fit for object sections located within this interesting region, then this wouldn't be the most significant interesting region. The Kuiper test is applied to all possible combinations of position and scale efficiently using a test region that expands and then shrinks as it moves to different positions.

\subsection{Creating objects from object sections}
\label{section:objects}
Once the {\sc CNHI} source finder has finished finding object sections in a chunk of the datacube, the object sections are combined into objects. Objects are created from object sections using a variant of Lutz one pass algorithm. The essence of Lutz one pass algorithm is to raster scan through an image or datacube building up the properties of every object as you go. Lutz crucial insight is that objects are simply connected so they pop out of the current image or datacube section being scanned (the scanline). This places a limit on the number of objects that need to be tracked and updated at any one time. Once the entire datacube is scanned, the rejection critierion is applied to the objects. The surviving objects are written to the output catalogue.

\begin{figure}[ht]
\begin{center}
\includegraphics[width=0.98\linewidth]{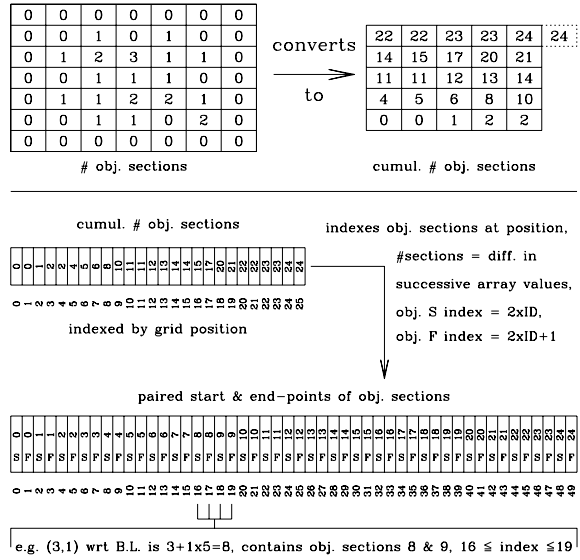}
\caption{An illustration of the novel sparse representation of 3D objects, that was developed for the {\sc CNHI} source finder.}
\label{fig:S3Drep}
\end{center}
\end{figure}

The crucial change to Lutz one pass algorithm is the use of sparse representations of 3-D objects. The use of sparse representations is what allows the {\sc CNHI} source finder to process the datacube in chunks. A new sparse representation, which consists of three components, was developed expressly for this purpose. The first component lists the RA and Dec widths of the object's bounding box. The second component lists the number of object sections that make up the object prior to a given RA, Dec position. The final component lists the channels that each object section begins and ends at. The first component indexes the second component, which then indexes the third component. The structure of this sparse representation is illustrated in Figure \ref{fig:S3Drep}.

The flaw of Lutz one pass algorithm is that it assumes objects are never encountered again after they `pop' out of the scanline. This assumption is quite easily invalidated when a datacube is too big to process in a single pass. A crude solution is to split the datacube into overlapping chunks, process them individually and then merge the detections within the chunk overlaps. This approach can easily produce erroneous results though, because it relies on the positions of a source's segments being sufficiently close to be merged together. A better method is to update each chunk's mask with all of the previously detected sources before processing them. To do this we need to be able to efficiently store the binary mask of every object in a readily accessible format. The sparse representation of 3-D objects presented here enables the {\sc CNHI} source finder to do exactly that. This approach to processing arbitrarily large datacubes also makes the {\sc CNHI} source finder readily amenable to distributed/parallel computing. 

An additional benefit of using the sparse representations is that the binary mask of every object can be written to the {\sc CNHI} output catalogue. Storing an object's binary mask allows us to run an arbitrary source parametrisation tool without having to re-run the source finder to generate the binary mask. 

The {\sc CNHI} source finder incorporates the sparse representation of 3-D objects in the following way. The flag array, which stores the object sections found in the datacube chunk currently being processed, is first updated for the objects found in previous chunks using their sparse representations. The {\sc CNHI} source finder then scans through the flag array until it finds a voxel that has been flagged as source. The software then searches through the flag array to find previously processed objects within the voxel's user-specified merging volume. This is a 3-D implementation of what is referred to as 8-connected linking in 2-D images. If there aren't any objects within the merging volume, then a new object and sparse representation is created. If a single existing object is found within the merging volume, then the source voxel is added to it. If multiple existing objects are found, then the existing objects are merged into a single object and the source voxel added to it. In each scenario the flag array is updated with the object id of the source that each source voxel belongs to. When existing objects pass out of the merging volume of all possible new objects, the user specified rejection criterion is applied to these existing objects. This ensures the minimal number of objects are stored in memory at any given time. The object ids of rejected objects and the memory used to store their properties and sparse representations are recycled. After the flag array has been scanned, the sparse representations and postage stamp images of the surviving objects are constructed/updated as required.

\section{Example results}
\label{section:examples}
An analysis of the completeness and reliability of the {\sc CNHI} source finder, and a comparison of its performance to other source finders is presented in \citet{Popping_2011}. This paper presents a complementary analysis to Popping et al. (2011), using the same point source (PS) and extended source (ES) test datacubes \citep{WP_2011}. The terms completeness, raw reliability, refined reliability, merging rate and fragmentation rate are defined and measured as in \citet{Popping_2011}. As in \citet{Popping_2011} and \citet{WP_2011}, the analysis presented here acknowledges the difference between raw reliability and refined reliability, the refined reliability being the reliability that is possible using post-processing to remove false detections from a source finder's output catalogue. From here on, the term `reliability' will refer to the raw reliability. Note that the refined reliability has a corresponding refined completeness, which accounts for the true detections that are incorrectly flagged as false detections during post-processing.

During visual inspection of the {\sc CNHI} source finder detections in various parameter spaces, it was discovered that the total intensity versus maximum voxel intensity (intensity of the source's brightest voxel) parameter space is the most efficient parameter space in which to post-process {\sc CNHI} detections. For a given object with a given total intensity, the more compact it is the brighter the maximum voxel intensity. For each threshold a simple cut (a line) in this parameter space was used to post-process the {\sc CNHI} detections. The line was adjusted for each threshold until the refined reliability was greater than 90\%. This required minimal effort, and is an example of the type of post-processing advocated in \citet{Serra_2011}. 

\begin{figure}[ht]
\begin{center}
\includegraphics[width=0.98\linewidth]{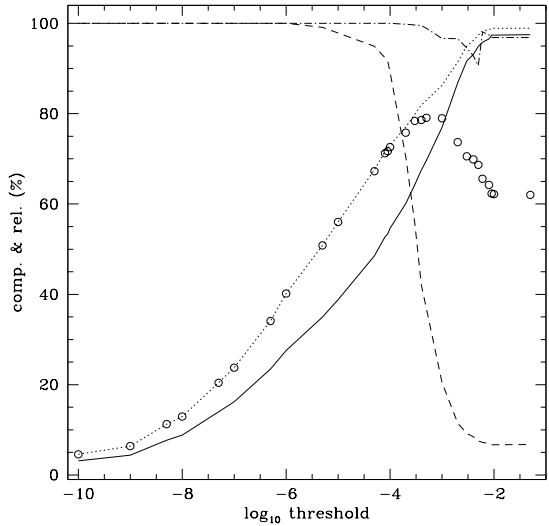}
\caption{Completeness (solid line), reliability (dashed line), corrected completeness (dotted line), refined corrected completeness (circles) and refined reliability (dot-dash line) curves for the PS datacube.}
\label{fig:CR_PS}
\end{center}
\end{figure}

\begin{figure}[ht]
\begin{center}
\includegraphics[width=0.98\linewidth]{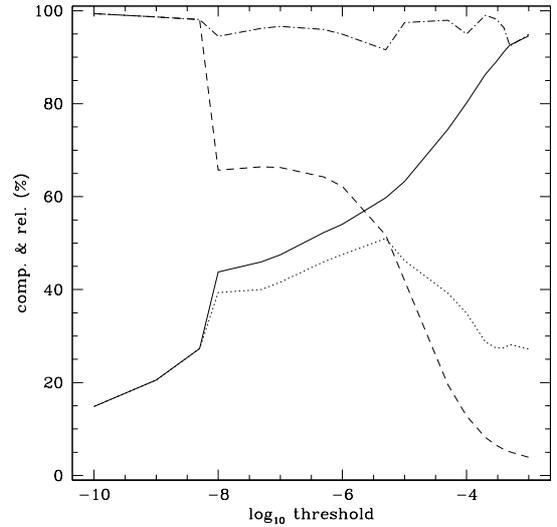}
\caption{Completeness (solid line), reliability (dashed line), refined completeness (dotted line) and refined reliability (dot-dash line) curves for the ES datacube.}
\label{fig:CR_ES}
\end{center}
\end{figure}

The completeness, reliability, refined completeness and refined reliability are plotted in Figures \ref{fig:CR_PS} and \ref{fig:CR_ES} for the PS and ES datacubes as a function of threshold. In Figure \ref{fig:CR_PS}, the inherent limitation of the {\sc CNHI} source finder is taken into account by measuring a `corrected completeness'. The corrected completeness is measured by excluding sources in the PS input catalogue with a full-width at half-maximum smaller than four channels. This corrected completeness therefore measures the completeness of the sources that the {\sc CNHI} is expected to find. The refined completeness in Figure \ref{fig:CR_PS} is that of the corrected completeness. The inherent limitation of the {\sc CNHI} is only taken into account for the PS datacube, because a significant fraction of the sources in the PS datacube have Gaussian profiles with FWHM's that extend over 3 channels or less. Uncertainties are not provided for the curves in Figures \ref{fig:CR_PS} and \ref{fig:CR_ES}, because to generate meaningful uncertainties requires generating many noise realisations. This is beyond the scope of this paper. It is also irrelevant, because the performance of the {\sc CNHI} source finder on the PS and ES datacubes is not a guarantee or guide to the performance of the {\sc CNHI} source finder on other datasets.

As expected, when generating the curves in Figures \ref{fig:CR_PS} and \ref{fig:CR_ES}, the choice of pre-threshold has as much of an impact on the completeness as the choice of threshold. To obtain the best completeness the pre-threshold should be set as high as compute resources and the datacube size will allow. Note that it was also observed that a pre-threshold larger than 0.1 or even 0.01 is excessive, and needlessly computationally expensive. Pre-thresholds larger than 0.1 or 0.01 are unlikely to improve upon the number of real sources that are recovered. For this reason a pre-threshold of 0.01 was used to generate Figures \ref{fig:CR_PS} and \ref{fig:CR_ES}.

The curves in Figures \ref{fig:CR_PS} and \ref{fig:CR_ES} illustrate that it is possible to achieve a good combination of refined completeness and refined reliability. For the PS dataset it was possible to achieve a refined, corrected completeness of $\sim 80\%$ with a refined reliability of $\sim 95\%$. A refined completeness of $\sim 50\%$ was achieved for the ES datacube, also with a refined reliability of $\sim 95\%$. For both datasets the curves in Figures \ref{fig:CR_PS} and \ref{fig:CR_ES} have negligible merging and fracture rates. 

Thresholds of $10^{-3}$ and $10^{-6}$ produce arguably the optimal combination of refined completeness and refined reliability in Figures \ref{fig:CR_PS} and \ref{fig:CR_ES}. For this reason, the performance of the {\sc CNHI} source when using these thresholds was examined further. First, the completeness was measured as a function of maximum voxel flux. Next, the fraction of total flux recovered by the {\sc CNHI} source finder was measured as a function of source total flux. Finally, the distribution of the difference between the true position of the sources and the position measured by the {\sc CNHI} source finder was determined.

\begin{figure}[ht]
\begin{center}
\includegraphics[angle=-90,width=0.98\linewidth]{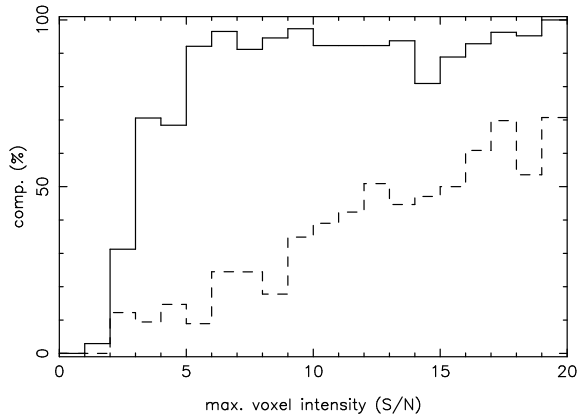}
\caption{The refined completeness as a function of source maximum voxel intensity. The refined completeness of the point sources (solid line) and extended sources (dashed line) is shown for thresholds of $10^{-3}$ and $10^{-6}$.}
\label{fig:CompMI}
\end{center}
\end{figure}

Figure \ref{fig:CompMI} shows that the {\sc CNHI} source finder can find almost all of the PS objects with a maximum voxel intensity five times brighter than the noise, with a high refined reliability. The {\sc CNHI} source finder also finds a significant fraction of PS objects with a maximum voxel intensity between three and five times the noise. Unfortunately, the {\sc CNHI} source finder does not perform as well at finding the ES objects. Visual inspection of the undetected ES objects revealed that they are `pancake' galaxies. These pancake galaxies are spatially resolved, but due to orientation only extend over a few channels. The bundling of \HI spectra acts as a crude spatial filter, and the bundling used here closely matches the spatial profile of the PS objects. This suggests that the {\sc CNHI} source finder can be improved by using multiple bundling schemes. This improvement would be equivalent to using matched filtering in three dimensions, with the spatial and frequency dimensions using independent filters.

\begin{figure*}[htbp]
\begin{center}
\includegraphics[angle=-90,width=0.32\linewidth]{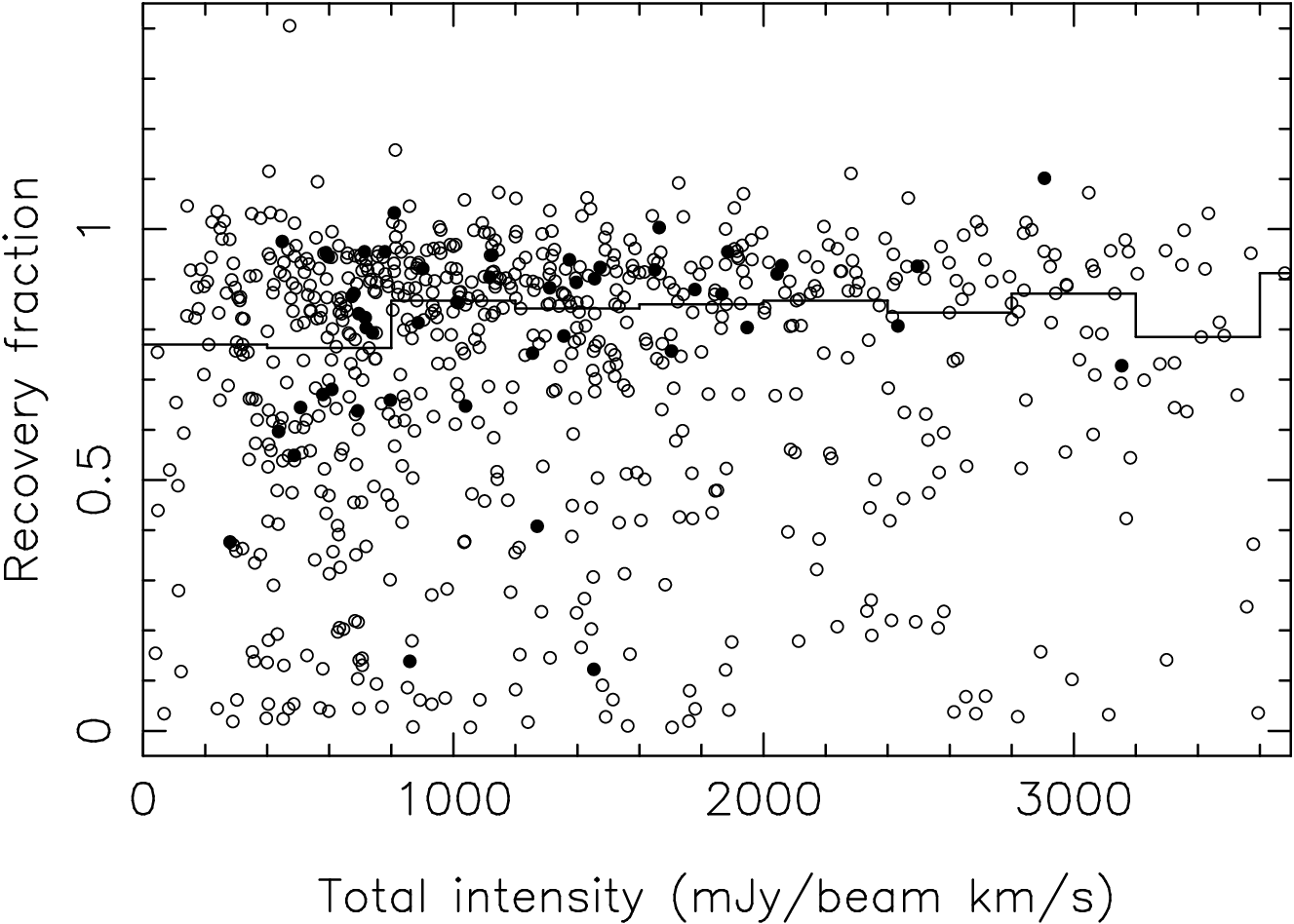} \includegraphics[angle=-90,width=0.32\linewidth]{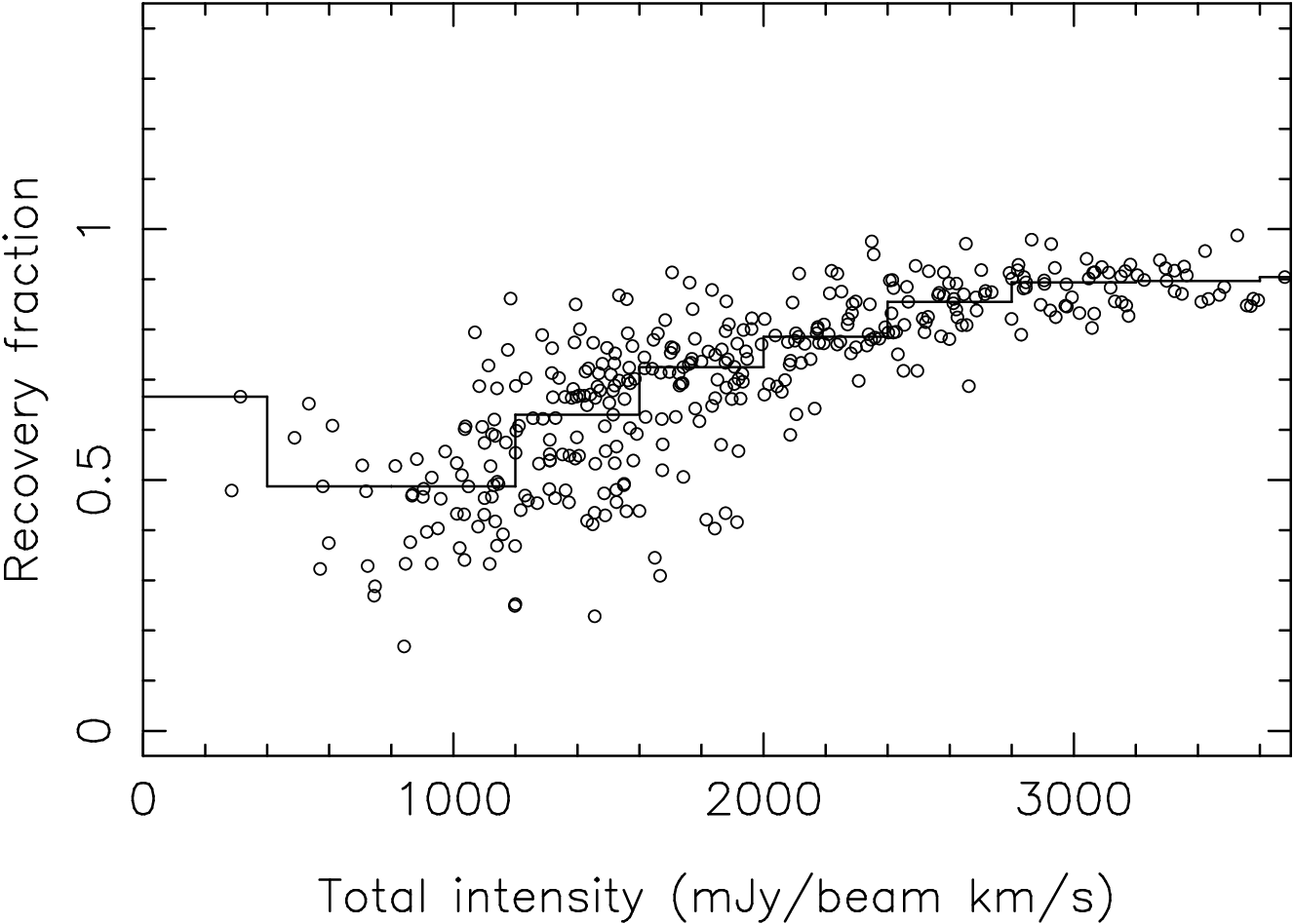} \includegraphics[angle=-90,width=0.32\linewidth]{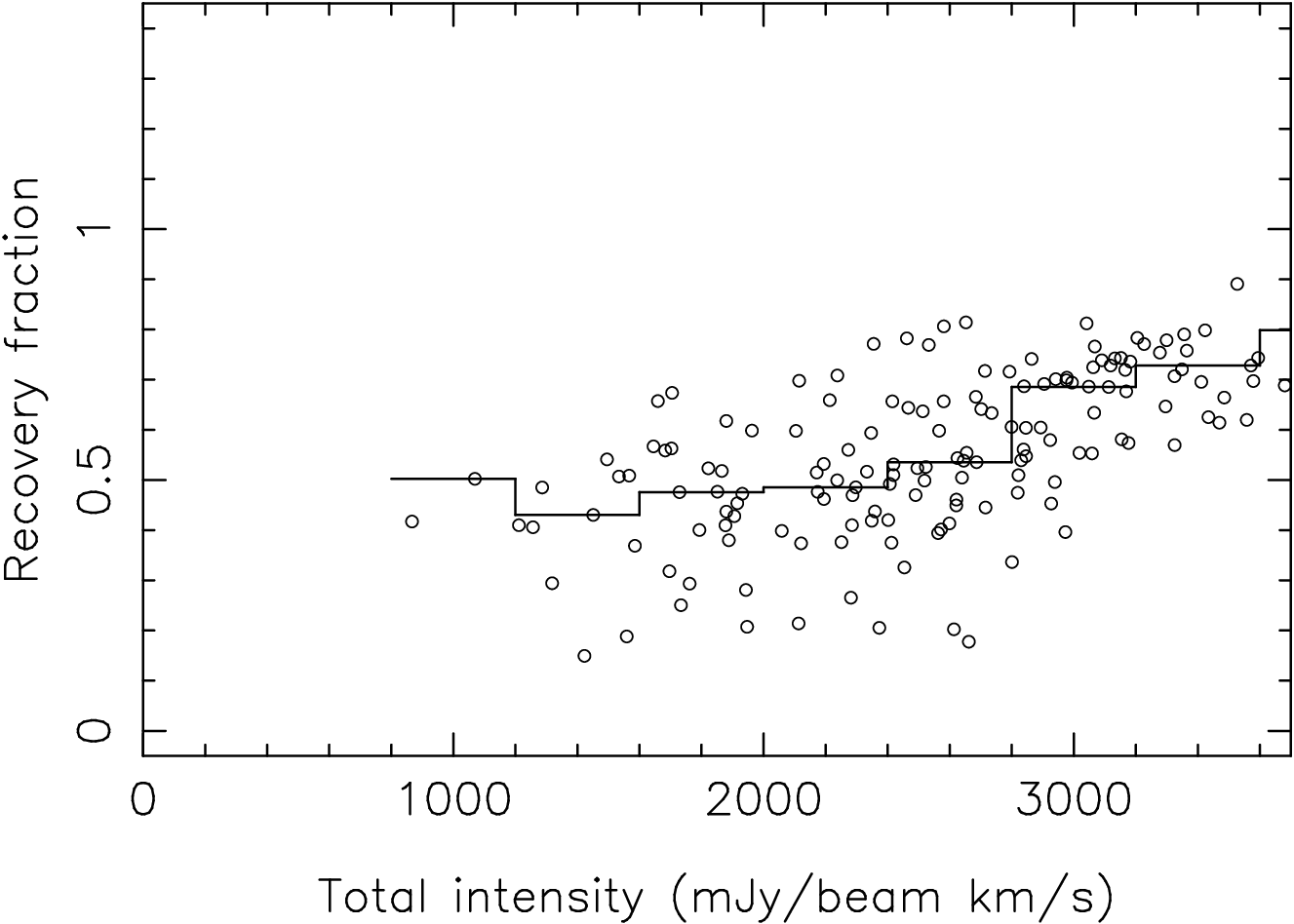}
\caption{The recovery fraction of the total intensity of sources in the PS datacube for thresholds of $10^{-3}$ (left), $10^{-5}$ (middle) and $10^{-7}$ (right). Sources that have been fragmented into multiple detections are represented using solid circles instead of hollow circles. The median recovery rate of the non-fragmented sources is overlaid as a solid line. The median is measured in bins of size 400 mJy/beam km/s. Note that solid circles are only plotted in the left panel. The appearance of solid circles in the middle and right panels is due to a high density of hollow circles.}
\label{fig:RF}
\end{center}
\end{figure*}

The fraction of each source's total intensity that is recovered by the source finder is plotted in Figure \ref{fig:RF}, and is referred to as the recovery fraction. The mean recovery fraction is measured after excluding fragmented sources, and is overlaid in Figure \ref{fig:RF}. A poor recovery fraction requires post-processing of each detection by a second tool to improve the source parametrisation. The mean recovery fraction for the $10^{-3}$ threshold in Figure \ref{fig:RF} asymptotes from $\sim 80\%$ to $\sim 90\%$ as sources become brighter. The recovery fractions greater than 100\% are a result of comparing total intensities measured in the noisy datacube to a reference total intensity measured in the noise-free datacube. This is the most meaningful comparison though, because it is sensitive to object masks that are either too large or too small. The recovery rate is as good as I would expect to do, without over-estimating the total flux. This demonstrates that minimal post-processing is required to improve the parametrisation of sources detected by the {\sc CNHI} source finder. 

Using more conservative thresholds of $10^{-5}$ and $10^{-7}$, which do not push as far into the noise as higher thresholds\footnote{The meaning of `aggressive' thresholds for the {\sc CNHI} source finder is the reverse of thresholds used in intensity threshold based source finders. For intensity threshold source finders, lower thresholds are more aggressive and push further into the noise.}, results in a median recovery fraction that asymptotes from $\sim 50\%$ to $\sim 90\%$ and $\sim 50\%$ to $\sim 70\%$ as sources become brighter. The median recovery fraction appears to asymptote more rapidly for more aggressive thresholds. This suggests that as the choice of threshold becomes more aggressive, the total flux measured for a source will converge to its true value. This suggests two things. First, the sources with low recovery fractions in the left panel of Figure \ref{fig:RF} are probably objects that have only just been detected, and using a more aggressive threshold would result in a higher recovery fraction. Second, an alternative use of the {\sc CNHI} source finder is as a source parametrisation tool rather than a source finder.

\begin{figure}[htbp]
\begin{center}
\includegraphics[angle=-90,width=0.98\linewidth]{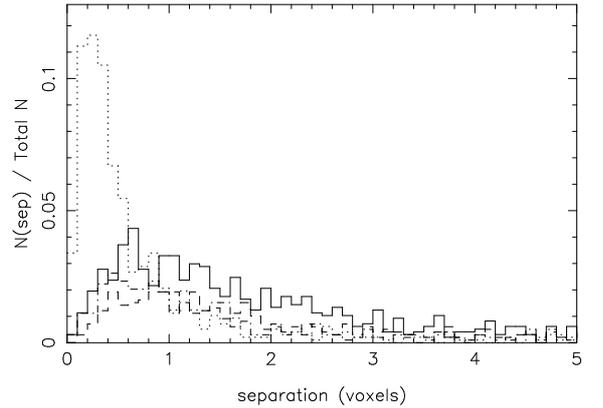}
\caption{The relative distribution of separations between the voxel centres and flux weighted voxel centres of sources and their corresponding {\sc CNHI} detections. The separations of the unweighted and intensity weighted PS positions is shown by the solid and dotted lines. The dashed and dot-dash lines show the separations of the unweighted and intensity weighted positions for the ES dataset.}
\label{fig:SD}
\end{center}
\end{figure}

The final component of the analysis is the distribution of the separation in voxels between the centre of the sources and their corresponding detection. The separation distributions of the voxel centres and intensity weighted voxel centres are presented in Figure \ref{fig:SD}. 

A separation between the centre of an object and its corresponding {\sc CNHI} detection reveals biassed source detection. The centres of an object and its corresponding detection are calculated as an unweighted mean of the voxel positions or a flux weighted mean of the voxel positions. It can be safely assumed that the voxel and weighted voxel centres of an object are accurate. A non-zero separation therefore arises due to a distorted source detection, where one region of the source is detected more than the rest. If the mean voxel centre separation is sharply peaked about a separation of zero, then we can conclude that {\sc CNHI} detections are on average unbiassed representations of the corresponding source. If the mean weighted voxel centre separation is similarly distributed, this demonstrates that the {\sc CNHI} detections are on average an unbiassed representation of the source's flux distribution. In Figure \ref{fig:SD} the unweighted (intensity weighted) separation distributions of both the PS and ES datasets are sharply peaked at 0.7 (0.3) and 0.9 (0.5) voxels. This demonstrates that the {\sc CNHI} source finder detections are typically a fair, unbiassed representation of the underlying source and its flux distribution.

\begin{figure}[htbp]
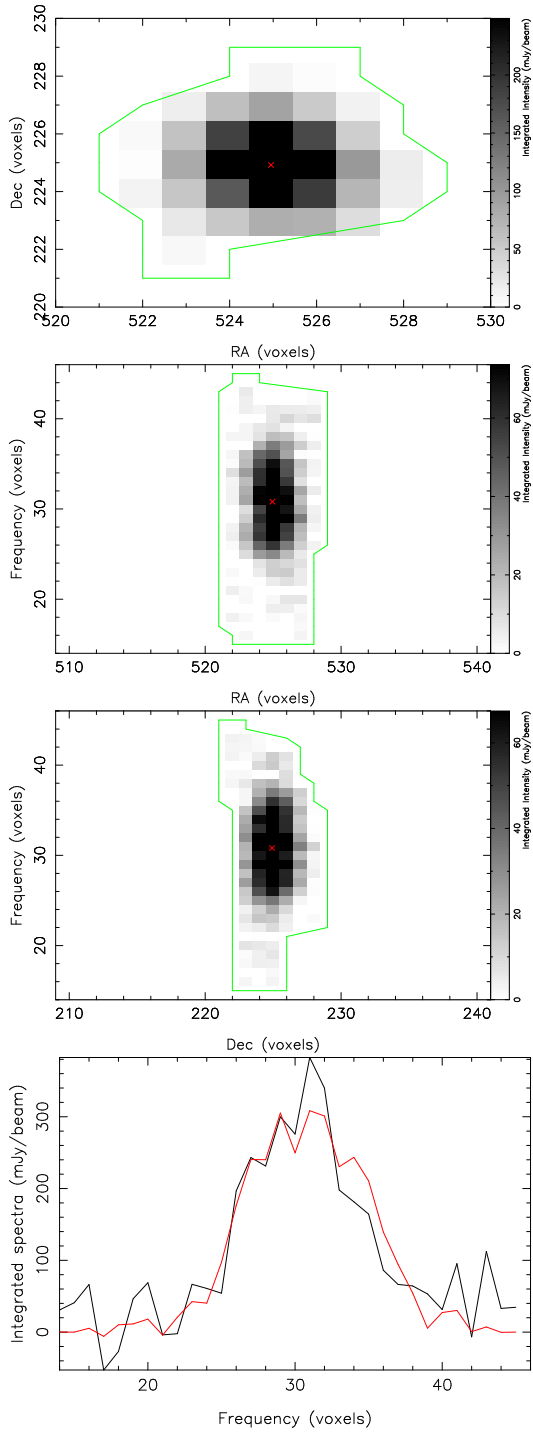

\begin{center}
\includegraphics[angle=-90,width=0.9\linewidth]{PS_demo_mom0.ps} \\
\includegraphics[angle=-90,width=0.9\linewidth]{PS_demo_RAPV.ps} \\
\includegraphics[angle=-90,width=0.9\linewidth]{PS_demo_DecPV.ps} \\
\includegraphics[angle=-90,width=0.9\linewidth]{PS_demo_spectra.ps}
\caption{Example postage stamp images produced by the {\sc CNHI} source finder for a PS object with a threshold of $10^{-3}$. The red cross marks the intensity weighted position of the object. The boundary of the object is marked by a green line. From the top the images are a moment-0 map, RA position-velocity diagram, Dec position-velocity diagram and integrated spectrum. The integrated spectrum consists of both the object (red line), and an integrated spectrum over the entire bounding box of the object (black line).}
\label{fig:PSdemo}
\end{center}
\end{figure}

\begin{figure}[htbp]
\begin{center}
\includegraphics[angle=-90,width=0.9\linewidth]{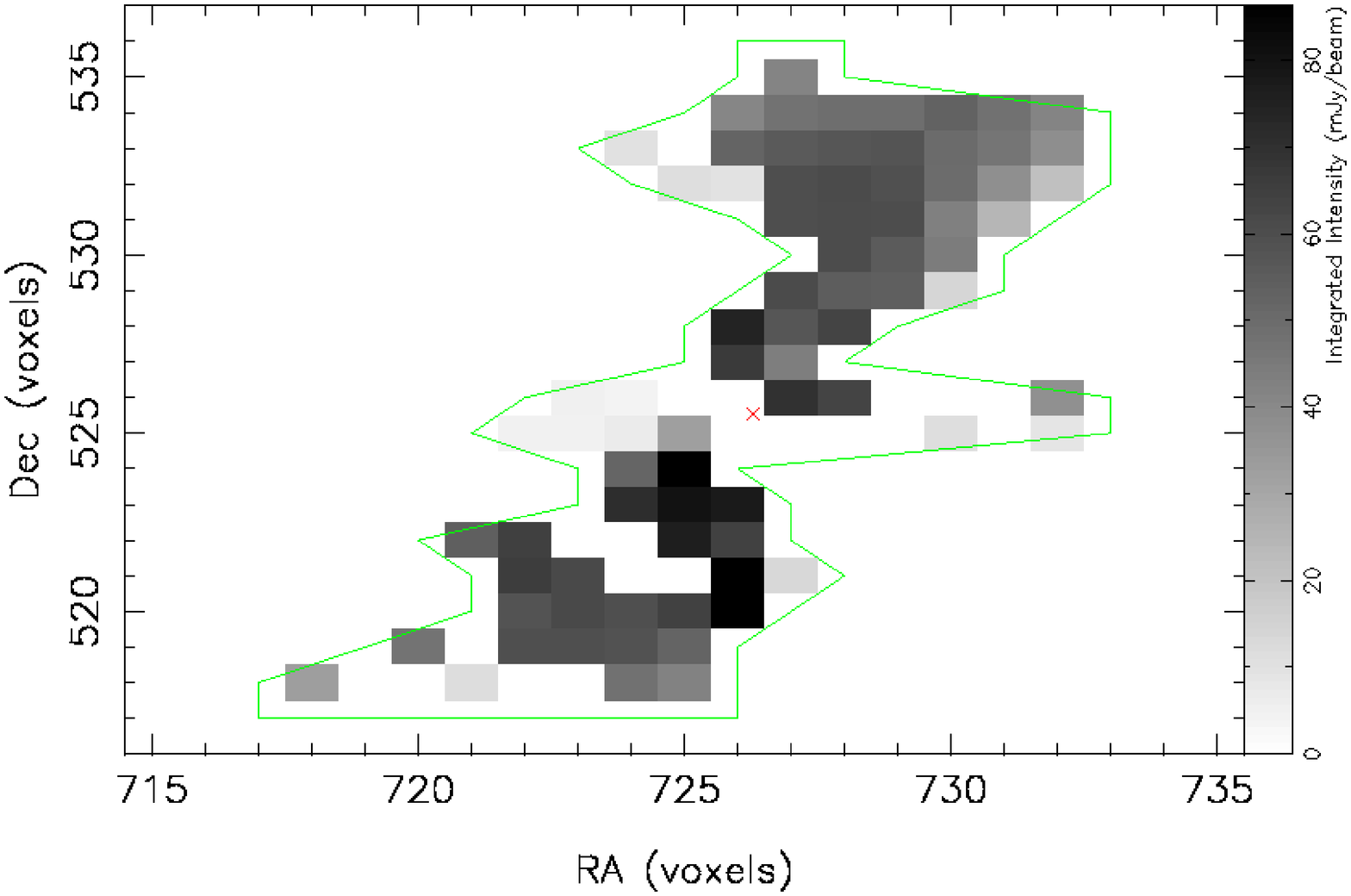} \\
\includegraphics[angle=-90,width=0.9\linewidth]{ES_demo_RAPV.ps} \\
\includegraphics[angle=-90,width=0.9\linewidth]{ES_demo_DecPV.ps} \\
\includegraphics[angle=-90,width=0.9\linewidth]{ES_demo_spectra.ps}
\caption{Example postage stamp images produced by the {\sc CNHI} source finder for an ES object with a threshold of $10^{-6}$. The intensity weighted position and boundary of the object are marked with a red cross and green line. The postage stamps are a moment-0 map (top), RA position-velocity diagram (second from top), Dec position-velocity diagram (second from bottom) and integrated spectrum (bottom). The red spectrum is the object and the black spectrum is a reference spectrum of the object's bounding box.}
\label{fig:ESdemo}
\end{center}
\end{figure}

This section finishes by presenting examples of the postage stamp images generated by the {\sc CNHI} source finder. Figures \ref{fig:PSdemo} and \ref{fig:ESdemo} are postage stamp images for two objects selected at random from the PS and ES dataset (one from each dataset). Both figures demonstrate the ability of the {\sc CNHI} source finder to find the boundaries of sources. Figure \ref{fig:ESdemo} also nicely illustrates that the {\sc CNHI} source finder is capable of detecting objects extending over many channels as a single object, rather than fragmenting it into two or more detections.

\section{Summary}
\label{section:summary}
WALLABY and other projects that will be carried out on next-generation radio telescopes ASKAP and MeerKAT herald the start of the data deluge era in radio astronomy. The sheer size of WALLABY datacubes necessitates automation of many tasks in the data reduction pipeline, that previously would have been carried out with some level of manual input by an astronomer. Complete automation of finding \HI galaxies in spectral datacubes is one of the challenges that is actively being investigated by WALLABY. 

The resolution and size of WALLABY observations poses a challenge for many existing automated source finders. This challenge arises from the underlying conceptual framework and algorithm that these source finders are based on, and not a flaw in the implementation. The {\sc CNHI} source finder has been developed using a conceptual framework that can handle the large size of WALLABY datacubes and takes advantage of the resolution. Treating a datacube as a set of spectra (akin to an IFU observation), it attempts to find sources by looking for regions in each spectrum that do not look like noise. This is achieved using a novel implementation of matched filtering. Instead of using multiple filters that describe various types of sources, a single filter describing the noise is used. Sources are detected using this noise filter by identifying regions that do not look like noise.

The performance of the {\sc CNHI} source finder was tested using the PS and ES datasets in \citet{WP_2011}. Analysis of the {\sc CNHI} source finder output demonstrated that a reasonable combination of completeness and refined reliability can be achieved. A refined completeness of $\sim 80\%$ and $\sim 50\%$ was achieved for the PS and ES datasets, respectively, with a refined reliability of $\sim 95\%$. The PS dataset is better than the $80\%$ completeness would suggest though, because the {\sc CNHI} source finder found $\sim 95\%$ of all PS objects with a maximum voxel flux $\geq 5\sigma$, with a refined reliability of $\sim 95\%$. This analysis also demonstrated that the {\sc CNHI} source finder recovers a significant fraction of the source flux. The recovery fraction asymptotes towards $100\%$ as the total flux increases. More aggressive thresholds (larger) result in a recovery fraction that asymptotes faster, and starts higher. This suggests an alternative use of the {\sc CNHI} source finder as a source parametrisation tool, that is used in tandem with another source finder. Finally, the performance analysis demonstrated that {\sc CNHI} detections of a source are an unbiassed representation of the source and its flux distribution. These results are very promising, and warrant further testing and refinement of the {\sc CNHI} source finder. 

There are three development goals for the {\sc CNHI} source finder. Further development of the {\sc CNHI} source finder will initially focus on incorporating multi-scale bundling. This will effectively achieve independently-scaled matched filtering in both the frequency and spatial dimensions. Additionally, the {\sc CNHI} source finder will have a simple intensity thresholding test added to it. Incorporating an intensity thresholding test will make the {\sc CNHI} source finder sensitive to sources occupying 3 or fewer channels. The final development goal is to incorporate fourier analysis, polynomial fitting and existing baseline structure removal techniques. Upon completing this next development cycle, the {\sc CNHI} source finder will be tested and tweaked using the next round of ASKAP simulations and the \HI Parkes All Sky Survey (HIPASS) \citep{2000ASPC..217...50S} datacubes. The HIPASS datacubes have been selected because they have a well defined source catalogue, contain a mixture of resolved and unresolved sources, have known artifacts and calibration issues and there is a potential for the {\sc CNHI} source finder to detect new sources.

\section*{Acknowledgments} 
The author would like to thank the rest of the WALLABY/DINGO source finding working group for many useful discussions and suggestions.

\end{document}